# Clarifying Cognitive Control and the Controllable Connectome

John D. Medaglia[1,2]

6/1/2018

**Abstract**

Cognitive control researchers aim to describe the processes that support adaptive cognition to achieve specific goals. Control theorists consider how to influence the state of systems to reach certain user-defined goals. In brain networks, some conceptual and lexical similarities between cognitive control and control theory offer appealing avenues for scientific discovery. However, these opportunities also come with the risk of conceptual confusion. Here, I suggest that each field of inquiry continues to produce novel and distinct insights. Then, I describe opportunities for synergistic research at the intersection of these subdisciplines with a critical stance that reduces the risk of conceptual confusion. Through this exercise, we can observe that both cognitive neuroscience and systems engineering have much to contribute to cognitive control research in human brain networks.

---

[1]Department of Psychology, Drexel University
[2]Department of Neurology, Perelman School of Medicine, University of Pennsylvania

**GRAPHICAL TABLE OF CONTENTS**

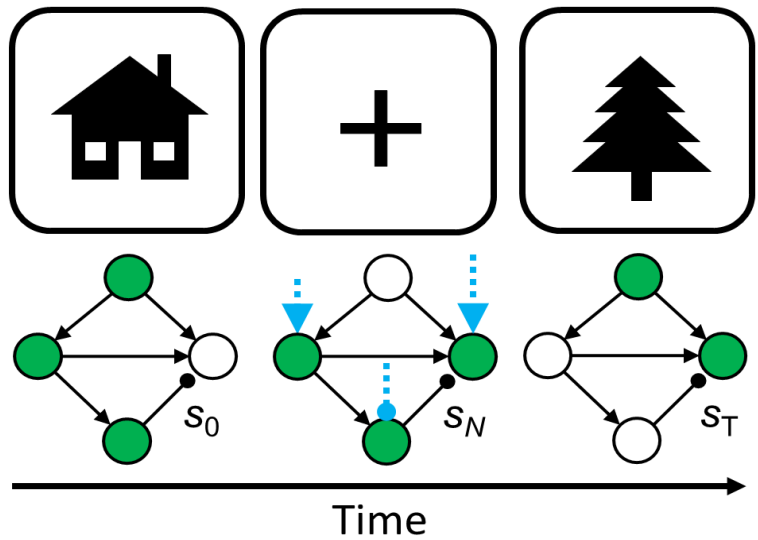

Figure: How can we carefully integrate cognitive control and systems engineering research for basic discovery and translation?

# INTRODUCTION

*Cognitive control* is one of several classes of cognitive processes that are central to healthy, autonomous human functioning. These functions help us to plan for our futures, inhibit distractions, and switch our thoughts and behaviors to achieve our goals. The brain circuitry that is responsible for cognitive control is becoming well-elucidated as cognitive neuroscientists endeavor to identify the regions responsible for specific functions and how those functions interact. The instruments and mathematical approaches available to us can provide unprecedented opportunity to observe and manipulate neural activity pertinent to cognitive control. This can facilitate discovery and potentially new opportunities for optimizing interventions for cognitive control deficits observed as a result of many etiologies, such as brain injury or neurodevelopmental disorders.

Meanwhile, approaches from a subdiscipline of systems engineering dating to the mid 1800s known as *control theory* allow us to describe, design and influence natural and artificial physical systems. Recently, some concepts from control theory have been applied to describe the organization of control in the human brain independently from the primary cognitive control literature. While these efforts are revealing interesting features of brain network organization and potential dynamic roles, the connections to cognitive control research may require several modifications to obtain a clear and consistent account. By clarifying cognitive control and control theory applied to human brain networks (or the "connectome" (Sporns, Tononi, & Kötter, 2005)), we can identify several exciting theoretical opportunities for researchers.

## Some words of caution

In human language, lexical similarity can lead to semantic ambiguity. The fact that the word "control" is used to refer to both of the topics at hand may promote enthusiasm, but we should be very careful. First, we should notice that the content and focus of the two fields emerged independently. Specifically, modern cognitive control research is an effort based in computational neuroscience to understand several processes thought to occur in the brain. In contrast, control theory more generally refers to system identification and how to manipulate dynamics in systems. This historical distinction means that there exist cultural, academic, and real-world differences regarding how the shared word – "control" – is defined, studied, and applied within these subdisciplines. It is very easy to confuse the meaning of the terminology shared between the fields, and practicing scientists typically have in depth training in only one field or the other. This will make communication between the fields challenging if practitioners from both sides do not take the time to understand what the other one means.

Second, we should notice that neither subdiscipline is "complete". There exists no perfect, universally endorsed account for exactly what cognitive control is, whether there are one or more forms of cognitive control, or how cognitive control is implemented in the brain. Moreover, there is no generalizable account for how to engineer a perfect control theoretic solution for goal in real-world networks, including brains. For the topic at hand, this should tell us that cognitive control research may benefit from outside influences. In addition, it provides engineers with a challenging context in which to refine concepts, procedures, and tools when attempting to apply control theory to a cognitive system.

Thus, to minimize confusion and promote effective collaborative work, I first briefly define "cognitive control" and "control theory" as separate classes of concepts with independent origins. Then, I describe emerging work on the "Controllable Connectome" as early success in integrating these concepts. Crucially, I focus on distinct questions that a control theoretic view is apt to contribute to cognitive control research and some early research in this area. Finally, I briefly introduce open questions from a control theoretic perspective to help us navigate the conceptual and practical issues that arise when using control theoretic concepts in real-world cognitive systems.

**COGNITIVE CONTROL**

Cognitive (or "executive") control (henceforth, CC) often refers to the functions that regulate and coordinate basic attention, memory, language, and action-related abilities to perform specific tasks (Botvinick & Braver, 2015). CC frequently refers to the ability to inhibit impulses or switch rules, actions, or the focus of attention (Davidson, Amso, Anderson, & Diamond, 2006). Sometimes, researchers also use CC to refer to working memory or behavioral performance monitoring (Ridderinkhof, Van Den Wildenberg, Segalowitz, & Carter, 2004). For our current purposes, we will consider CC to be the human cognitive ability to "configure itself for the performance of specific tasks through appropriate adjustments in perceptual selection, response biasing, and the on-line maintenance of contextual information" (Botvinick, Braver, Barch, Carter, & Cohen, 2001). Thus, CC mediates between goals and action, attuning our perceptions and behavior to meet task demands.

We can understand the role of CC in cognition based on its recruitment, modulation, and disengagement (Botvinick et al., 2001). CC is often cued by the attempt to perform a difficult (and often novel) task, and is thought to be recruited as a function of the *expected value* of the control (Shenhav, Botvinick, & Cohen, 2013). Online CC is modulated in response to variations in performance, suggested by decreases in speed and increase in accuracy following on-task errors (Botvinick et al., 2001) and the statistical distribution of control-demanding trials within a task (Botvinick et al., 2001, Lindsay & Jacoby, 1994, Kane & Engle, 2003). Finally, CC demands often decrease with task practice as processes required to perform the task become more automatic (Logan, 1989), leading to the disengagement of CC (Botvinick et al., 2001).

In the human brain, the specific representation of CC is not completely resolved, but some major conclusions are supported at a high level (i.e., brain regions) of brain organization. The dorsal ACC (dACC) may monitor conflict (Kerns et al., 2004) and cue the recruitment of CC by integrating information about rewards and costs expected from a control-demanding task (Shenhav et al., 2013, Shenhav, Cohen, & Botvinick, 2016). Once cued, CC involves mechanisms in the prefrontal cortex that inhibit, switch, retrieve, and select representations and processes throughout the brain (Aron, Robbins, & Poldrack, 2014, Domenech & Koechlin, 2015, Duverne & Koechlin, 2017, MacDonald, Cohen, Stenger, & Carter, 2000), which interact heavily with parietal attention-mediating mechanisms (Yantis et al., 2002, Bisley & Goldberg, 2010). Collectively, these regions can heuristically be considered to be a "core" of the computational mechanisms involved in CC. Of great importance in both CC research and control theory, CC is commonly investigated using measurable behavioral variables. These variables represent the "output" of CC from the brain through the body, and will provide the anchor that helps us integrate CC and control theory: the brain has internal neural states responsible for CC, and behavior provides the ultimate signals of CC as a person interacts with the world (see Fig. 1).

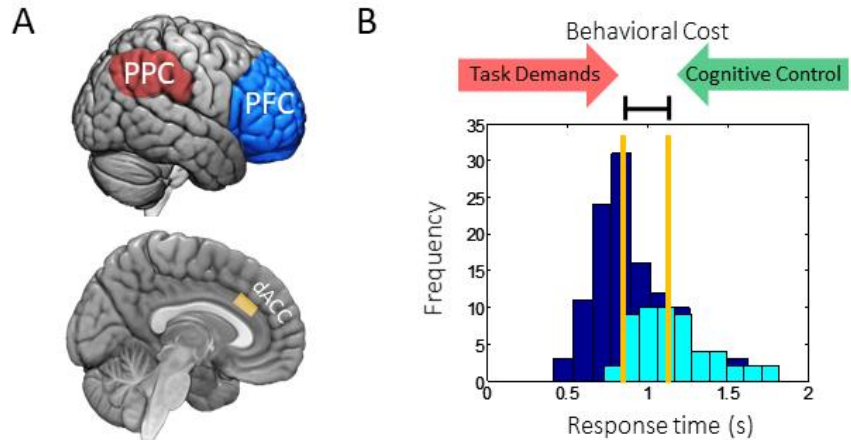

Figure 1: **The Brain and Behavior in Cognitive Control Research.** *(A)* CC is recruited as a function of the expected value of its engagement computed by the dACC, control influences from the prefrontal cortex (PFC) and attention switching mechanisms in the posterior parietal cortex (PPC). *(B)* Behavioral costs of CC are often computed as the difference (e.g., in median differences across types of trials, or differences on a specific trial) in response times between a "control-demanding" experimental condition (teal bars) and a reference condition (blue bars). For example, this difference could represent a control-demanding "interference effect" during a Stroop task or "switch cost" during a set-switching task. The behavioral cost is thought to result from the push-pull between increasing task demands offset by recruited cognitive control (Botvinick et al., 2001). Accordingly, we can anticipate that influencing the brain's processing of task demands or core cognitive control circuitry will change the measured behavioral costs.

Notably, there are diverse accounts of role of the ACC in CC. For example, the ACC is triggered by unpredictable environments and events, and the ACC's response potentially enhances learning and emotion-enhanced memories via noradrenaline response mediated by the locus coeruleus (Verguts & Notebaert, 2009). Further, the ACC has been suggested to participate in response monitoring via the prediction of response outcomes (PRO). Specifically, one component of the model predicts the likelihood of the various previously experienced outcomes, each of which is associated with a positive or negative valence. The second component of the model compares the actual and expected outcomes, producing a signal when a discrepancy occurs (Alexander & Brown, 2010). One view aims to reconcile neuropsychological theories focused on energizing and motivating behavior at a prolonged timescale with reinforcement and cognitive control models that operate at a trial-to-trial level. In this account, ACC supports the selection and execution of coherent behaviors over extended periods of time using a heirarchical reinforcement learning model (Holroyd & Yeung, 2012). Finally, one account maintains that the ACC encodes value-related information more generally, including the average value of exploring alternative choices and the degree to which internal models of the environment and task must be updated. These value signals result from a history of recent rewards integrated over multiple time scales, and do not necessarily represent conflict or difficulty signals (Kolling et al., 2016).

**CONTROL THEORY**

While CC research emerged in the mid 20th century, control theory's origin is often attributed to James Clerk Maxwell's description of the operation of "governors" in the 19th century (Maxwell, 1868). At the time, centrifugal governors were often used to keep the velocity of a machine (such as a steam engine) nearly uniform despite variations in driving power or resistance. Maxwell focused on *stability*. He observed that a machine's motion consists of a uniform motion plus a disturbance that resulted from several component motions. Per Maxwell, a good governor was that which, along with the machine it controls, exhibited a continually diminishing disturbance or an oscillating disturbance with a continually decreasing amplitude. A "genuine governor" was one that applied both proportional and integral control actions – that is, the error between the state of the system and its desired state at a given time, as well as how long and far the measured process has been from the desired state over time. Later, Nicolas Minorsky would contribute a theoretical analysis for the *derivative* of the error - its current rate of change. Minorsky's contribution, initially rejected by naval operators due to resistance from personnel, supported the later emergence of modern proportional-integral-derivative (PID) controllers (Minorsky, 1922).

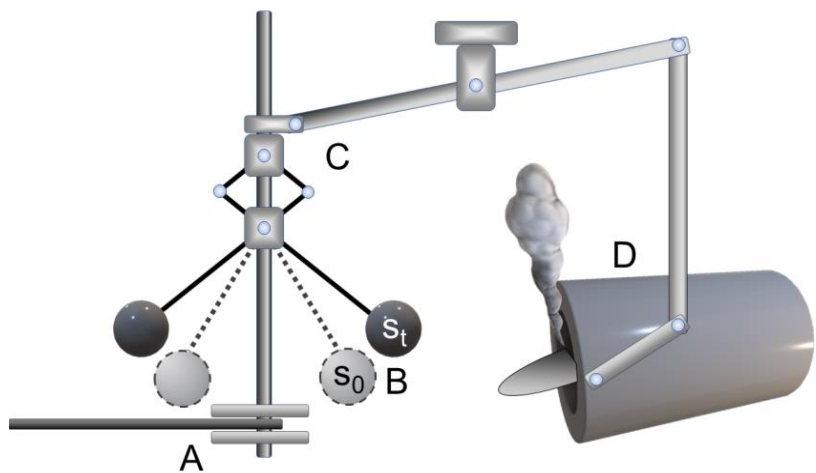

Figure 2: **A schematic Watt governor.** James Watt introduced an early flyball governor in 1788 to control the velocity of steam engines. It contained both a sensor and a control mechanism. *(A)* A connecting belt from the engine spun a cylindrical shaft. *(B)* As the engine's speed increased, the centrifugal force created by the shaft's rotation spun the flyballs. *(C)* As the flyballs elevated from a resting state $s_0$ to a higher state $s_t$ with increasing engine velocity, a mechanism transferred the flyball displacement along a stem connected to a steam valve. *(D)* As engine speed increased, a steam valve opened, reducing engine pressure and velocity.

For our current purpose, following Maxwell's classic focus on governors, we can identify some of the essentials of control theory using an example early governor (See Fig. 2). In this governor, the flyballs act as a *sensor* that provided *feedback* about the speed of the machine. The flyball governor mechanism serves as a *controller*, which mediates energy transfer into a system. Intrinsic in this device is a computation of the *error* between the system state and its desired state (the current *versus* the target velocity). Because of the relationship between the input to the governor and its output, engineers can exploit the relationship to achieve the desired control *goal* of the machine's velocity. Observing this, Maxwell gave us a more general mathematical basis for enacting control in mechanical devices.

Following from this example in an early, simple control device, we can notice that control theory is more generally useful whenever feedback can be predicted or measured in a system, and whenever we can use a mechanism to influence the state of a system. We must be able to observe a system in order to prospectively control a system; indeed, our theoretical success in designing a control system is fundamentally limited by its *observability*, or the measure of how well a system's internal states can be inferred based on direct measurement or the system's outputs (Ogata & Yang, 2009). In contrast, *controllability* denotes the theoretical ability to move a system around the entire space of its possible configurations under specific constraints (Ogata & Yang, 2009). Together, observability and controllability form dual aspects of the same problem. Once we have established adequate observability of a phenomenon, we can consider robust approaches to identify control strategies to influence that phenomenon.

In control theory, we can examine the *state controllability* and the *output controllability* of a system. State controllability refers to the ability for an external input to move a system from an initial condition to a final condition in a finite time. Output controllability is the same notion but with regards to the final output of the system. Further, state controllability typically refers to the possibility of transferring the state between two values, while output controllability refers to the possibility of forcing the output along a desired controllability. State controllability is typically less restrictive than output controllability because the relationship between the number of controllers and states is not necessarily constrained, whereas in output controllability the number of outputs is constrained to be at maximum the number of controllers.

Notably, a state controllable system is not necessarily output controllable if it demonstrates feedforward restrictions to the output mechanism (i.e., some internal states cannot influence the output states). In addition, an output controllable system is not necessarily state controllable. For instance, if a state space is much larger than the output space, there can be a set of states associated with each output (see Fig. 3). This latter observation implies a notion of output *redundancy*: as long as at least one reachable state exists that can produce a given output, the output of the system is possible, and potentially output controllable. This type of redundancy is distinct from but related to the control theoretic notion of *robustness*. Specifically, a robust control system is one that resists perturbation (often in the form of uncertainty in some parameters (Dorf & Bishop, 2011)). In this example, redundancy could imply robustness because some variation in the parameters of the system may achieve the output state via multiple paths.

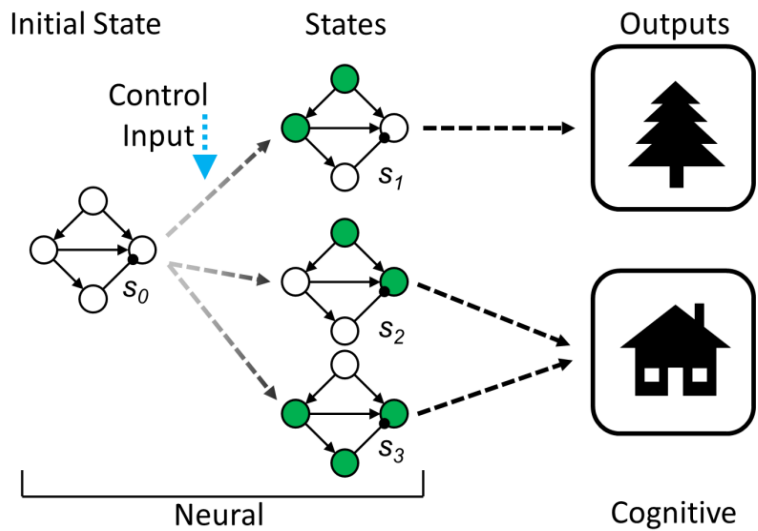

Figure 3: **Neural states and outputs in cognition.** In an arbitrary perceptual example, we can imagine an initial state $s_0$ of a set of neurons that can exist in several other states ($s_1, s_2, s_3$) that result when the system is influenced by a *Control Input* (e.g., cognitive control or exogenous stimulation, including environmental cues or invasive or noninvasive neuromodulation such as via brain stimulation). However, only $s_1$ is associated with a tree, whereas $s_2$ and $s_3$ are associated with a house. If a person perceives a house and we can't directly measure the internal (neural) state, we can't know which state the system is in. However, we may be able to identify an input that results in the perception of a house, and the perception of the house is more "robust" because, all else being equal, if the system cannot achieve $s_3$, then $s_2$ is still a viable state to achieve to ensure that the person will perceive the house.

**Two classes of control theory**

In his analysis of governors, Maxwell made another key contribution with consequences that still influence modern control theory. The first mathematical analysis of instability was performed by George Biddell Airy, who attempted to use differential equations to understand governor instability. The equation could not easily be made linear, and he could not reach a general conclusion (Airy, 1840). Maxwell achieved limited success in linearizing equations of motion obtained in his work studying Saturn's ring's (Maxwell, 1859), and extended this work to his study of governors (Maxwell, 1868). Maxwell had successfully applied the idea that linearizing differential equations around a desired operating point facilitates more straightforward mathematical solutions. Here, we should briefly distinguish linear and nonlinear control.

**Linear control theory**

Linear control theory is applicable to devices that obey a superposition principle, meaning that the output of a system is proportional to its input. Linear systems can be described by linear differential equations. Linear time-invariant (LTI) systems, which have parameters that do not change with time, form a major subclass of linear systems. LTI systems are amenable to analysis with many frequency domain-based mathematical techniques, providing control solutions for many systems of interest. While truly linear systems are exceedingly few in the real world, the study of linear systems provides a basis for the study and control of simple devices, such as flyball governors and toilet bowl float valves.

**Nonlinear control theory**

*Most real systems are nonlinear.* This means that they do not obey the superposition principle: their outputs are not proportional to their inputs. These systems are typically described by nonlinear differential equations. In general, the small classes of mathematical techniques developed to handle them are more difficult and apply to narrow categories of systems. Because of this difficulty, nonlinear systems are commonly approximated by linear equations, which works well to a degree of accuracy for small ranges of input values. However, linearization comes at the additional cost of masking phenomena such as chaos: nonlinear systems can exhibit behavior that appears random due to exquisite sensitivity to initial conditions (Kellert, 1994). In addition, even if a system is completely predictable by solving its differential equations, stochastic real-world systems that include noise may exhibit unpredictable behavior. For example, small changes in weather in one location produce complex, unpredictable effects across space and time, which is why very long-term forecasting is impossible with current techniques (Shukla, 1998, Selvam, 2017).

**Single-input-single-output (SISO) and multi-input-multi-output (MIMO) control systems**

Another important distinction in control theory is between SISO and MIMO control systems, which are at two ends of systems that can also include single-input-multiple-output (SIMO) and multi-input-single-output (MISO) systems. SISO systems are generally less complicated because their dynamics are restricted to those that mediate between a single influence and a single response. Examples include automotive cruise control, wireless communications with a single antenna and single receiver, and audio systems that receive an audio signal and output sound waves from a speaker. MIMO systems can encompass all of the dynamics of an SISO system, plus the interactions that occur when multiple inputs and outputs are intertwined within a system. Many computer and mechanical systems are MIMO, involving numerous input channels, mediating devices that associated or dissociate inputs, and several output channels. In the human brain, single neurons can be considered MIMO systems if all the dendritic processes and neurotransmitters are considered, or MISO systems if only post-synaptic potentials are considered, such as in "integrate-and-fire" neural models (Burkitt, 2006). In addition, neural network models are widely used as part of the control mechanisms in SISO and MIMO systems (Chen, Ge, & How, 2010). At the psychological scale, we can consider many cognitive systems to be MIMO systems because they receive inputs from and send outputs to many other systems in the brain.

**How can we apply control theory to cognition?**

So, what do these control theoretic concepts have to do with neurons and cognition? In simplest terms, we can view cognitive control as one of many potential "reverse engineering" problems in cognitive neuroscience. We can aim to understand how CC is produced by a specific system – the brain. Crucially, to the control theorist, CC results in measurable behaviors (outputs) as the result of internal states (neural activity). In principle, the control engineer could aim to integrate feedback about the output information with control signals to influence cognitive control – such as with targeted pharmacology or brain stimulation – to achieve healthy, desirable cognitive-behavioral states. Her success will be related to the fidelity of our models of CC and access to technical and theoretical control solutions. The engineer's mentality provides us with some simple objectives to keep in mind for the road ahead (See Fig. 4).

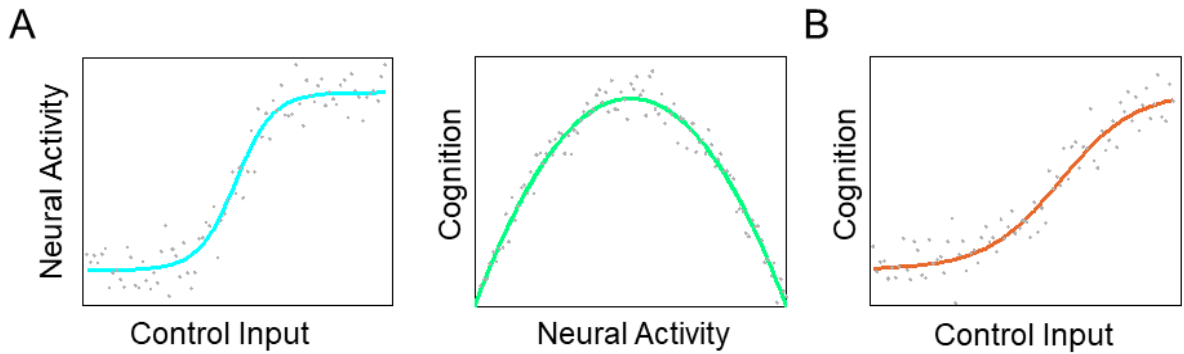

Figure 4: **Goals for control theory in cognitive control research.** In principle, we can focus on two relationships when linking control theory to cognitive control research. Schematically, *(A,left)* we should consider the relationship between control inputs and resulting neural activity. *(A,right)* We should also identify the relationships between neural activity and cognition, often measured using behavioral variables in cognitive control research – the lynchpin of cognitive neuroscience. Because neural activity joins the panels in *(A)*, the *system identification* aspect of control theory will involve a focus on identifying equations that express neural dynamics' natural and manipulable relationships with cognitive control behavioral output. *(B)* Notably, we can also be interested in effects directly from inputs to behavioral outputs in control theory. In each graph, the characteristic function will be associated with a noise profile, which limits our ability to uncover and exploit the basic input-output curves.

There have been some notable influences from control theory on biology and psychology. Early in the 20th century, Walter Cannon observed that many physiological systems maintain homeostasis by stabilizing certain variables with respect to changing environments using similar processes to those used in control theory (Cannon, 1929). Cybernetics broadly concerns the science of communication and control in organisms and artifical devices, and has made deep and broad uses of control theoretic concepts since its early days (Ashby, 1961). In the 1970s, control theory involving loops between perception and behavior was offered as an alternative to both behaviorism and psychoanalysis (Powers & Powers, 1973). Further, control theory has had a deep influence on motor control systems theory (Gallivan, Logan, Wolpert, & Flanagan, 2016, Wei & Körding, 2010, Scott, 2004, Cisek & Kalaska, 2010). In this tradition, we will consider some emerging progress specifically concerning CC and control theory.

**Cognitive control in the controllable connectome**

To combine CC research and control theory, several distinct challenges confront us. By identifying these themes, we can further refine goals for researchers interested in closing the substantial gap between formal control theory and CC science. First, we will consider major open questions. Then, approaches and evidence from an emerging line of research that merges cognitive network neuroscience and control theory via *network control theory* are summarized with special attention to the promise and limitations of work to date.

**Framing questions within a control theoretic view of the brain and mind**

We are now well-prepared to clarify some key distinctions and commonalities between control theoretic and CC concepts. From a control theoretic perspective, we can consider various *scales* of control: we can view the brain in its entirety or its subparts as control mechanisms. Alternatively, we can view the brain as a system to be controlled, such as through behavioral, pharmacological, or brain stimulation interventions. Following the intuition that the brain can be both "controller" and "controllee", we can consider what and how to target the brain for what purposes in CC research.

**The brain as controller and controllee**

We can sensibly view the brain both as a controller and controllee. Neurons and their surrounding neurophysiology influence one another. These physical influences are assumed to produce cognitive functions via information processing (Churchland & Sejnowski, 2016, Dayan & Abbott, 2001). To the control theorist, these assumptions meet the necessary and sufficient conditions to start her work: there is physical system in which we can identify characteristic interactions among elements that result in observable outputs, which are typically neural and ultimately cognitive. Thus, identifying neural systems as *controllers* is an exercise in *system identification* in engineering (Ljung, 1998): identifying the entities and activities responsible for phenomena (Illari & Williamson, 2012) and expressing them in concise (if counterintuitive or complicated (Dayan & Abbott, 2001, Markram, 2006)) mathematical formalisms. In addition, we can view any use of an exogenous influence on the brain as an external control input. We should expect that the effects of the stimulation will depend on the design of the input and the characteristics of the system. An illustration can guide this basic intuition (Fig. 5).

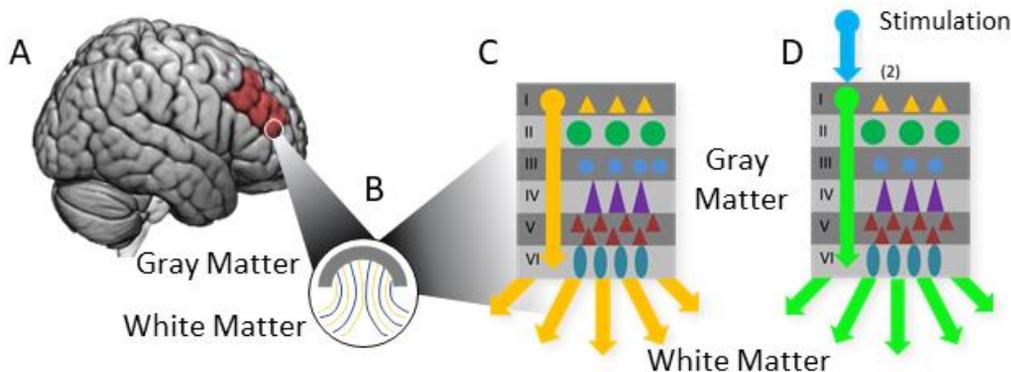

Figure 5: **A schematic for the natural "Controller/Controllee" duality in neural control engineering at an arbitrary neural scale.** *(A)* At any given location in the brain, such as the prefrontal cortex *(A)*, we can consider the local organization of *(B)* white matter (which comprises the anatomical connections between brain regions) and gray matter (which comprises the neural cell bodies which sum neural inputs and initiate outputs). Some proportion of white matter connections relay signals into the gray matter (yellow connections) and some relay signals out of the gray matter (blue). *(C)* At a smaller scale, gray matter is generally arranged into layers that include numerous different types of cells organized within layers (represented with colored symbols) which communicate among each other. Certain layers ultimately project information out of the local layers to other regions, representing the net outputs of computations that occur within gray matter that will influence other brain regions. In this case, the pictured brain region is the "controller" that influences other regions. For simplicity, the interlayer interactions are not pictured here, but crucial to the computations in real networks. *(D)* Exogenous brain stimulation can be applied to local brain regions (now the "controllee"), modulating the typical neural activity across layers (here represented by the green arrows, and ultimately the output of the brain to other regions.

Viewing the brain as both "controller" and "controllee" reinforces an intuition that cognitive neuroscientists have exploited for decades: manipulating a system can provide strong inferences about the function of the system. In this regard, the control theoretic view of the brain is nothing new. Indeed, control theory's consistency with our prevailing intuitions about experimental manipulation in cognitive neuroscience is one of its strengths. What control theory offers is rigor and opportunities in CC research that are not necessarily obvious in classical CC research by itself.

The challenge for the control theorist in CC research is at once simple and daunting. It is simple to casually observe that a CC system is a special case of a control system. It is daunting to consider how to obtain a complete system identification for a CC system in the brain. Generally, system identification in engineering is difficult and often yields models that lack parsimony or have poor prediction power (Bohlin, 2013). Strategies for system identification should involve careful consideration of the identification experiment, the set of models that might provide a suitable description of the system, the criteria for selecting the "best" model that describes the data, and whether the final model is "good enough" for the problem at hand.(Ljung, 1998). These questions define a sequence of questions that follow from one another in the process of system identification (Bohlin, 2013), and will be central to control theoretic approaches to CC. Specifically, we ultimately should desire to find models that have predictive validity, are parsimonious, and accurately reflect the configuration of a system and the time-varying states that support CC.

Considering this challenge, how can we proceed? Here, control theory helps tune our intuitions based on our goals and clear parallels to control theoretic ideas and terminology. We can align CC research with classical control theoretic practices to our advantage. For many years, CC research has focused on system identification via approaches that range from more abstract cognitive modeling – where the details of neural implementation are not necessarily modeled – to models that encode details of neural implementation (O'Reilly, Herd, & Pauli, 2010, Botvinick & Cohen, 2014). Notably, neurally-focused models of cognitive control do not represent all neural details presumed to contribute to the system. This provides a crucial foundation for control theory: the appropriate level of neural precision required to describe psychological functions could ostensibly do away with certain details. This would allow us to further hone our control theoretic approaches and questions. These independent developments in cognitive and neural modeling indicate that control theory was ripe for a more formal integration with cognitive neuroscience.

In a general sense, we can consider neurons to be small controllers that can be organized into larger control systems (Kawato et al., 1987; Liu et al., 2016; Mnih et al., 2015; Narendra & Parthasaranthy, 1990). However, cognitive neuroscience allows us to attribute more specific psychological functions to these systems in biological neural networks. It is crucial to notice that CC systems are *psychological* control systems. The CC systems observed in humans are sophisticated, naturally occurring, robustly adaptive control systems for guiding behavior and cognition (Carver & Scheier, 2012) when selecting among prepotent responses (Ridderinkhof et al., 2004). This high-level control is executed through considerable neural computation where distinct neurons encode goal-related information (e.g., spatial direction (Sarel et al., 2017), goal value (McNamee et al., 2013). In this context, the psychological system represents goals and executes processes to achieve them, and the brain systems that execute CC are naturally occurring instances of neural control engineering solutions to these problems.

Thus, in the case of CC, we find concepts that are familiar to the control theorist. A person has a behavioral goal to achieve. CC operates as part of a "closed loop" in a broader neural system that maintains goals. CC is cued when there is a mismatch ("error") between desired performance and actual performance (Alexander & Brown, 2010; Shenhav et al., 2013; Shenhav, Cohen, & Botvinick, 2016) and influences processes and representations across other networks. CC dynamically fluctuates within and between tasks to keep the person on task and performing as best as possible. To synthesize this notion further, the CC system is a "controller" that influences the rest of the system to adaptively maintain low error between actual and desired performance.

At present, what cognitive neuroscience at large could gain from control theory is access to powerful computational techniques that facilitate rigorous system identification (Garnier, Gilson, & Laurain, 2009, Ljung, 1992, Ugalde, Carmona, Reyes-Reyes, Alvarado, & Mantilla, 2015) to map the relationship between stimulation to neural function, neural-to-neural function, and neural function and cognitive-behavioral output. These tools can be applied to identify parsimonious, robust models of input-output relationships for any combination of components we may focus on. Once the system is expressed with a valid mathematical model, the model's predictions can be applied and tested iteratively. This latter intuition is often applied by experimental neuroscientists and control theorists alike (Frank & Badre, 2015). However, the success of our efforts in applied neuroscience may be enhanced enriching our conversation about CC with concepts and techniques used in formal control theory.

We can expect that any system identification will be successful to the extent that we acquire sufficient and valid data to detect true relationships. The scale and fidelity of our measurements will limit any given system identification, and in turn will limit our ability to represent or accurately control a neurocognitive process (e.g., for clinical gains). The neuron and researcher thus share a common limitation: just as one neuron has limited senses and influences in the context of its neural network, the researcher is limited by her behavioral and imaging measurement techniques and the potency and precision of any given manipulation using input. As control theorists, we can consciously aim to improve our measurements and control devices to describe and influence CC systems. As is the case for much of human cognition, there are significant limits to acquiring adequate input data for system estimation and testing. Neural measurement techniques vary greatly in terms of invasiveness and spatiotemporal precision that increasingly allow us to study the basis of CC more thoroughly, but there is no technique that allows us to completely measure behavioral inputs and neural states *in vivo*. Fortunately, the *output* data are arguably more measurable because cognitive scientists have created numerous paradigms that elicit measurements of cognitive control, such as inhibition and switch costs and selection and retrieval demands. Thus, the reverse engineering challenge for control theory in CC will eventually require highly spatiotemporally resolved techniques that allow for a rich system identification that encodes the appropriate environmental inputs and neural state dynamics.

Fortunately, control theory suggests a useful parallel for neural systems and our experimental approaches in CC research. A "good enough" control scheme (Motter, 2015, Medaglia et al., 2017) is one that achieves a goal despite technical and computational limitations. This involves identifying control solutions that interact locally with the natural dynamics of the system in noisy conditions with limited observability and control constraints (Motter, 2015). Although this is a generally anticipated design challenge for all control systems, it suggests something deeper for CC researchers: natural CC systems must be evolutionarily selected solutions to solve cognitive-behavioral control problems under constraints on physical energy, controllability, and observability, with further requirements for CC systems to be both robust and flexible. Energy, robustness, observability, controllability, and flexibility can all be grounded in statistical physics and engineering theory, which facilitates novel analyses and the opportunity to discover fundamental dynamics. In kind, when aiming to control CC systems, the wisdom of the applied neuroscientist who aims to control neurocircuits is that she needs only to identify a "good enough" solution to reach a goal, conceding *a priori* that a perfect one is by definition unattainable.

**What should we study and control?**

Well-oriented to the general challenges, what should we study as control theorists interested in CC? Unequivocally, the brain. Control theory forces us to confront the inadequacies of abstract models by orienting us to the basic representation, activation, and manipulability of CC through physical neural influences. Robust system identification approaches should be applied to calibrate us to the natural relationships among exogenous, neural, and cognitive variables, with full due attention to the data sampling requirements for good model identification (See Fig. 6; Ljung, 1998).

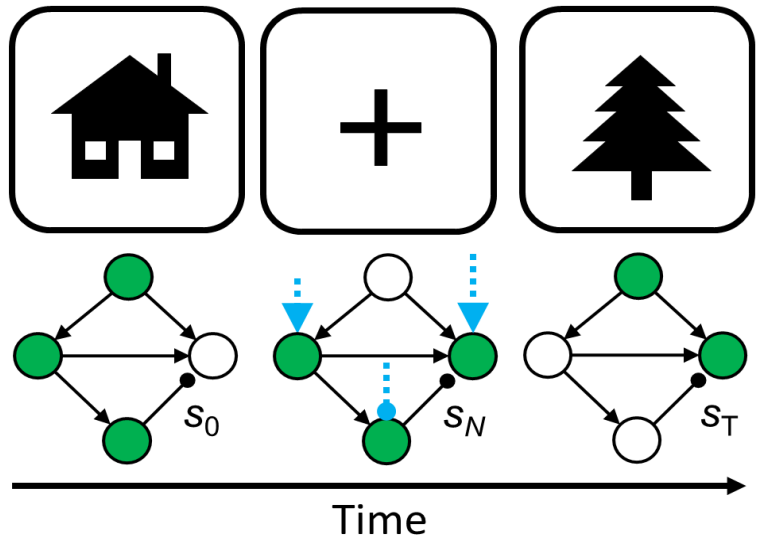

Figure 6: **A basic question in cognitive control research.** From this toy example, we can observe that a fundamental control problem in an attention switch is to move the state of a perceptual system from a state visually encoding a house *(left)* to a state encoding a tree *(right)*, such as with a switch of attention associated with an oculomotor saccade. The blue arrows represent the control process, which influences the relevant perception-action network from one state to the other. The blue arrows could be naturally occuring cognitive control, cognitive control moderated by exogenous stimulation, or stimulation alone. The details of the controller is what we aim to "reverse engineer" from this initial intuition.

At the heart of CC research we should be concerned with resolving issues of complexity (Bassett & Gazzaniga, 2011) including nonlinearity (Glanz, 1997, Motter, 2015), chaos (Schiff et al., 1994) hierarchy (Badre, 2008, Betzel & Bassett, 2016), and representation and process pairs (Braver, 2012). To investigate these phenomena thoroughly enough to claim we have truly discovered mechanisms of CC – that is to say, mechanisms so clearly articulated that we can demonstrably reproduce them (Medaglia, Lynall, & Bassett, 2015) – we require measurement, neural manipulations, and robust model identification strategies with adequate observability.

We should be concerned with facts about how computations in one region are robustly and flexibly (enough) communicated to another to relay CC signals under specific goal constraints. Because CC signals are fundamentally neurophysiological phenomena, we should identify the relevant signals within noise, why and to what extent they are robust, and how they exploit the controllability of perceptual and action circuitry across the brain. We should scale our neural measurements and manipulations to examine for influences that begin in one scale and influence others. Concepts from control theory suggests that opportunities to use exogenous stimulation to understand cognitive function have been sub-optimally utilized. This is likely because of historical divides between the fields. The basic intuitions of observability and controllability in control theory, experimentation to find optimal control strategies, and model-fitting can be combined to find natural contingencies among exogenously induced brain responses and behavior in a rigorous, parametric fashion. In this effort, neuroscientists and engineers at all scales using various methods of investigation are welcome to help identify actionable models of CC and its responses to exogeneous influences.

**Evidence from control theory applied to brain networks**

Some initial efforts to study cognitive control from a control theoretic perspective have used concepts developed in *network control theory* (henceforth, NCT) (Liu, Slotine, & Barabási, 2011, Ruths & Ruths, 2014). NCT concerns the control of processes in *networks*, which are systems that consists of observable, discrete elements ("nodes") and the connections among them ("edges"). The brain is a special case of a network, which has led to great enthusiasm for the utility of formalisms from *network science* (Lewis, 2011), often based in *graph theory* (West et al., 2001), to provide a formal mathematical basis with which to identify structures and processes that support cognition (Medaglia et al., 2015, Sporns, 2014, Bassett & Sporns, 2017). Building on this intuition and parallel developments from NCT, it is possible to ask questions about controllability in complex brain networks (Gu et al., 2015).

Just like nodes in brain networks have different roles (Rubinov & Sporns, 2010), NCT allows us to identify nodes' controllability within a network under some assumptions about the dynamics within the network. A node's controllability describes its ability to drive the network into any arbitrary desired state when injected with an input. For example, in networks with discrete time, time-invariant linear dynamics, analytic solutions can be used to identify the *modal*, *boundary* and *average* controllability of nodes within the network (Pasqualetti, Zampieri, & Bullo, 2014). These describe a node's role in easily driving the brain to a a hard to reach state, easily driving the brain into integrated or segregative activity among brain nodes, and easily driving the brain to an easy to reach state. "Easy" and "hard" refer to the energy demands required to move among states on an "energy landscape", which defines a system's state transitions and associated energy costs (see Fig. 7).

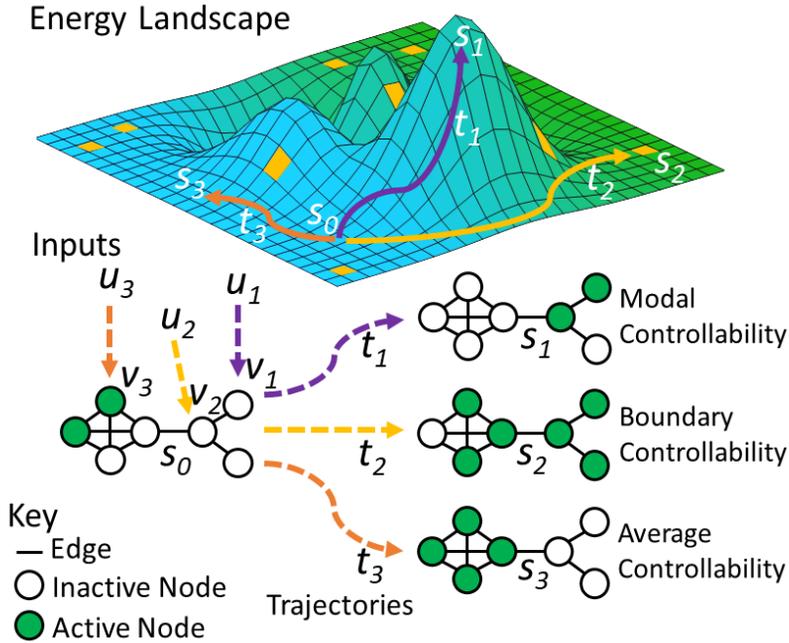

Figure 7: **Network controllability and energy landscapes.** *(Top)*: A schematic energy landscape where each cell represents a possible state $s_N$ of the network. Peaks represent states with high energy. Moving from cell to cell constitutes a *trajectory* ($t_N$) along the landscape. Moving farther or up an incline indicates a higher cost trajectory. Yellow cells represent special states where activity is integrated or segregated among all possible states. *(Bottom)* in response to an input $u_N$ to nodes in different positions on the network given an arbitrary initial state $s_0$, the network can be driven along distinct trajectories. Input $u_1$ to node $v_1$ will drive the brain into distant, hard to reach states (e.g., $s_1$ via trajectory $t_1$) and is thus high in modal controllability. Input $u_2$ to node $v_2$ will drive the brain into integrated or segregated states across clusters (e.g., $s_2$ via trajectory $t_2$), and is thus high in boundary controllability. Input $u_3$ to node $v_3$ will drive the brain along trajectory $t_3$ into nearby, easy to reach states (e.g., $s_3$ via trajectory $t_3$), and is thus high in average controllability. The relationship between a neural state and its associated cognitive representation(s) and process(es) must be empirically determined at the appropriate neural scales. We can study how the brain naturally deploys CC to navigate the landscape as well as how to design control inputs to guide the mind.

Gu and colleagues examined the controllability of the brain in the first application of NCT to macroscale white matter networks estimated with diffusion imaging (Gu et al., 2015) (see Fig. 8 for a methodological schematic). They found that commonly identified networks, including the cingulo-opercular and fronto-parietal CC networks, exhibited dissociable patterns of strong controllers. Specifically, they identified that modal, boundary, and average controllers were represented in fronto-parietal, cingulo-opercular, and so-called "default mode" systems, respectively. Speculatively, this suggests that at least some of the cognitive role of each system is attributable to how its regions are positioned to drive the brain along prototypical dynamic trajectories. These distinct subnetwork control profiles may have an intuitive parallel to the broader observations in cognitive neuroscience. In particular, the fronto-parietal network is frequently engaged in demanding, high-conflict, novel tasks (Cole et al., 2013, Dosenbach et al., 2007), the cingulo-opercular network is persistently engaged during online task performance, potentially mediating inter-network communication (Power & Petersen, 2013), and the default mode network may regulate transitions among many commonly accessible states across cognitive tasks (Raichle, 2015). Importantly, even under idealized dynamic assumptions, the brain is theoretically exceedingly difficult to control through any single region (Gu et al., 2015).

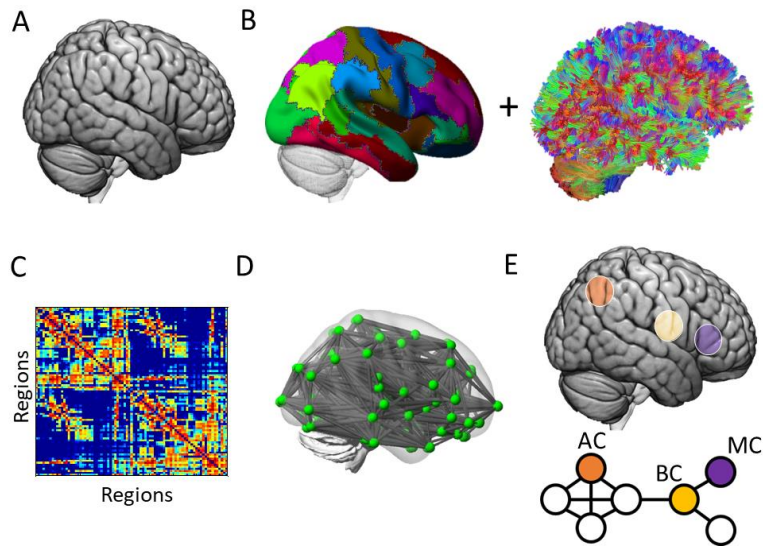

Figure 8: **Node controllability at a high level in diffusion imaging.** Network controllability analysis can identify the putative control roles of individual nodes in a network, such as has been applied to networks constructed from diffusion weighted imaging. Schematically, *(A)* within the brain, *(B, left)* anatomical regions can be represented in a parcellation combined with an estimate of the white matter pathways (e.g., "streamlines", fractional anisotropy) of the brain reconstructed from diffusion imaging *(B, right)*. *(C)* The streamlines can be used to construct a weighted region × region graph, or representation of the number of streamlines. *(D)* The location and identity of the regions and streamlines is retained in the original brain volume, facilitating *(E)* node controllability analysis to identify regions's characteristic control roles. AC = average controller, BC = boundary controller, MC = modal controller. Note that each node will exhibit some degree of controllability; here, maximally high ranking and distinct controllers are schematically represented.

Building from this work, Betzel and colleagues used an linear controllability analysis and found that hub regions in the brain's "rich club" (Van Den Heuvel & Sporns, 2011) participate in many states that minimize state transition energy, and that communicability between brain regions predicts which brain regions compensate when input to another is suppressed (Betzel, Gu, Medaglia, Pasqualetti, & Bassett, 2016). Moreover, across adolescents, the brain becomes more optimized for control whereby the average and modal controllability of brain regions is higher in later adolescence (Tang et al., 2017), facilitating asynchronous communication across the brain. Importantly, a shift in controllability away from neocortical regions toward subcortical regions was shown to be negatively related to a general cognitive efficiency score, suggesting that dominant controllability in the neocortex facilitates higher-ordered cognition (Tang et al., 2017); however, the specificity of relationships to individual functions was not examined. In addition, the position of executive control and attention systems was found to be well-positioned to affect a range of optimal brain state transitions and exhibit less specificity in control processes following mild brain injury (Gu et al., 2017). Over neurodevelopment, network controllability was associated with differences in executive function performance between males and females (Cornblath et al., 2018).

**The Path Ahead**

Studies to date represent a small subset of the potentially infinite questions to ask at the intersection of CC and control theory research. Notably, they all involve linear, time-invariant discrete time dynamics and measurements of discrete connections ("edges") among discrete neural components at a high level of organization. Moreover, the roles of nodes within the brain are typically assumed to have uniform dynamics. In addition, each application of NCT typically does not contain information consistent with computational theories of CC. Specifically, the notion of "control" in the brain is represented only by the general NCT definition with simple dynamics, not the formal cognitive sense. Put differently, *a priori*, any particular region or neuron's controllability is a feature of only its position in the topology of the system, not a feature of its position in addition to its internal computations and time-varying dynamics. Fortunately, each of these limitations can be further investigated within control theory, whether or not formalisms from linear NCT are used.

**Integrating Computationalism and Control Theory**

CC research has a strong foundation in computational neuroscience. Recall that the brain's initial deployment of CC can be predicted by the expected value of control (Shenhav et al., 2016). When CC is deployed, we can anticipate that a cascade of neural events occur. Coarsely, these events may compute error (Garavan, Ross, Murphy, Roche, & Stein, 2002), select and retrieve relevant representations (Altmann & Gray, 2008), adapt to feedback (Botvinick et al., 2001) and otherwise initiate processes across the brain to achieve goals (Dolan & Dayan, 2013). These events occur across spatiotemporal scales within the brain and represented by diverse neural dynamics. Some events, such as error and value computation, may occur in relatively discrete anatomical regions such as the dACC, but communicate heavily with cortical and subcortical systems through intricate anatomical pathways (Shenhav et al., 2013, 2016). Other events may be more appropriately thought to be the sequential and parallel activation of distributed perceptual and motor programs under directed attention (Zacks, Speer, Swallow, Braver, & Reynolds, 2007).

The computational processes involved in CC were identified independently from any formal notion of network controllability. However, these processes form part of the system identification, where CC systems are themselves viewed as control systems within the brain. Imporantly, models of CC are conceptually distinct from recent work applying NCT. Specifically, CC models do not necessarily encode details of the white matter organization in the brain, they typically emphasize the roles of a few brain regions, and the roles of individual regions are assumed to be computationally distinct (i.e., the regions have their own internal states). Thus, if we are interested in a satisfying account of CC as a brain network control process, several tensions should be reconciled. Specifically, we should be concerned with the scale of control, the computational and dynamic diversity among regions in the brain, and the positions (i.e., topological roles) of CC regions within the larger controllable connectome.

As findings from NCT suggest, it is possible that certain local (i.e., regional) computations are positioned at certain critical "control points" in the network to govern high-level dynamics. In this case, we would predict that key CC regions have control roles that guide specific dynamics to achieve goals. These dynamic roles could be to retrieve and select representations or processes. To integrate these perspectives, we can note that NCT provides an analytic basis for defining the control roles independently from current models of CC. Then, we can test for a CC region or system's network control role, such as that identified in (Gu et al., 2015, 2017), and –critically– for a relationship between the node's controllability and cognition (Tang et al., 2017). For example, identifying that CC-relevant regions exhibit task-related activation that is related to network controllability or the point of origin for network-level changes would be an important step in connecting NCT to CC systems. Next, in addition to a CC region's position on the network, we should examine regions' specific computations as roles within a formal control system that can be represented by a system of differential equations. This should involve controllability analyses both for states involved in CC and outputs resulting from the deployment of CC. This would also reveal potential CC architecture robustness and adaptability to damage (Gu et al., 2017). Crucially, as part of our analysis, we must eventually associate specific sets of states with specific cognitive representations and processes to define coherent control targets for CC systems, and for control theoretic approaches designed to influence CC.

One explicit example is the use of a proportional-integral-derivative (PID) control in modeling adaptive behavior. Specifically, the proportional-integral component of a PID model was found to be an effective model for participants' performance in noisy and changing environments (Ritz, Nassar, Frank, & Shenhav, 2017). In particular, proportional-integral control predicted behavior in both discrete transitions and gradual transitions, relying less on history-dependent control in the former case. Importantly, the findings were consistent with and extended beyond the standard delta-rule model of error-driven learning (Nassar, Wilson, Heasly, & Gold, 2010). These results provide evidence that the basic principles at work in one control theoretic mechanism widely applied in control theory can also describe some aspects of human behavior.

This research indicates that some well-known control theoretic principles from the 1800s may have been discovered by natural selection and become active in natural neural systems. More broadly, this could mean that there is a broader space of mathematically defined control schemes that explain controlled cognition and other adaptive functions in human cognition. Pragmatic approaches to determining input-output relationships in systems with robust mathematical tools suggest that control theory can help us find models that may not otherwise have occurred to us.

**Nonlinearity and Noise**

Because the brain is perhaps the quintessential nonlinear complex network, we should develop nonlinear models in CC research. This should further include the contributions of noise to the system's dynamics. While linear models used in NCT support inferences that can be validated in nonlinear demonstrations (Muldoon et al., 2016, Tiberi, Favaretto, Innocenti, Bassett, & Pasqualetti, 2017), the relationships that characterize neural interactions are diversely nonlinear, and this diversity drives much of the complex dynamics observed in real neural systems (Deco, Jirsa, & McIntosh, 2011). In addition, neural systems are noisy (McDonnell & Ward, 2011), and neural noise can drive important dynamic features such as information processing (McDonnell & Ward, 2011) and multistability (i.e., multiple stable states in neural networks (Deco et al., 2011). Fortunately, addressing nonlinear phenomena in noisy systems that exhibit multistability is a well-known problem in control engineering (Pisarchik & Feudel, 2014). Geometric approaches are emerging to determine how to drive multistable systems from one attractor to another (Wang et al., 2016). The challenge for the CC scientist is to identify the advantages for a particular nonlinear representation of the system in describing how CC operates and how to reliably influence CC with exogenous input. Finally, recall that another important feature of real, nonlinear neural systems is that they exhibit chaotic behavior (Schiff et al., 1994, Korn & Faure, 2003). Thus, robust control solutions for CC systems as controller and controllee will ultimately include "good enough" differential equation models that encode noise, exhibit realistic feedback and multistability, and chaotic behavior across empirically observable parameter ranges within CC systems.

**Translation**

Throughout this article, the relevance of control theory to *translational* issues in CC research have been implicit. For a translational researcher, the goal is to obtain a good enough solution to a clinical problem. Here we can see an obvious strength to thinking like a control theorist. If a person's CC function is impaired, we should identify practical solutions to improving it. As a control theorist knows, the better our models of CC, the better we may be able to identify attractive solutions. Importantly, because our models are never perfect, the good-enough philosophy in control theory means that we can identify our constraints on influencing an impaired CC system (e.g., limited dynamic models and access to neurons), and do our best to "reverse engineer" a solution within the constraints. To this end, we need only focus on good experimental methodology and robust system identification approaches to model basic input-output relationships among our interventions, the brain, and behaviour (recall Fig. **Error! Reference source not found.**).

There are a few specific contexts where control theoretic inquiry in CC research could become clinically relevant in the near term. With respect to emerging work using network control theory to identify dissociable roles for regions in brain networks (Gu et al., 2015, 2017, Tang et al., 2017, Betzel et al., 2016), network controllability measures could be paired with noninvasive or invasive brain stimulation to test whether topological control roles are associated with changes in CC function. If these control roles differentially predict changes in CC performance in response to stimulation, we will find evidence for a mechanistic role for theoretical controllability in the brain. We should examine the sensitivity and specificity of any such findings to enrich our understanding about the role of controllable brain topology in CC *versus* other cognitive functions.

In addition, broader interest in closed-loop paradigms such as those used in deep brain stimulation for Parkinsonism can be designed to be superior interventions to open-loop strategies (Rosin et al., 2011). More broadly, technological capacities are making closed-loop approaches increasingly available and interesting to applied neuroscientists (Zrenner, Belardinelli, Müller-Dahlhaus, & Ziemann, 2016) and psychiatric groups (Widge et al., 2017). Closed-loop applications are now appearing in transcranial magnetic stimulation (Kraus et al., 2016, Siebner, 2017, Karabanov, Thielscher, & Siebner, 2016). The promise for CC dysfunction is unexplored, but the principles are readily available. Given the role of deep brain structures in cortico-striatal CC processing, strategies originally designed to ameliorate motor systems at the cost of some CC functions could be investigated with control optimization techniques that attempt to reduce motor symptoms with constraints on adversely affecting cognitive control. In some cases of profound head injury or epileptic syndromes, techniques innovated for seizure control (Berényi, Belluscio, Mao, & Buzsáki, 2012) and obsessive-compulsive disorder (McLaughlin, Stewart, & Greenberg, 2016) could be adapted for use on neocortical systems thought to be responsible for CC. Fortunately, because many neocortical regions involved in CC are accessible with noninvasive stimulation techniques, success from open-loop brain stimulation strategies in noninvasive brain stimulation (Brunoni & Vanderhasselt, 2014, Hasan, Strube, Palm, & Wobrock, 2016, Rubio et al., 2016, Summers, Kang, & Cauraugh, 2016), we should be optimistic that pairing real-time adaptive noninvasive brain monitoring and stimulation techniques in closed-loop designs could improve our ability to intervene in CC and more generally on cognition in the long run.

**CONCLUSIONS**

The reader is now prepared with a high-level overview of the conceptual distinctions and interdisciplinary opportunities at the intersection between cognitive control and control theoretic research. It is clear that while these fields developed independently, human cognitive control systems are an evolutionarily sculpted and adaptable control system in human brain networks. This makes them amenable to robust reverse engineering and system identification approaches. Once sufficient numerical solutions for dynamics are identified, we can examine the relationship between the neural trajectories that represent cognitive control, which enact control on cognitive systems across the brain. Given good enough information, we can apply these models for potential clinical gains. This forms the frontier of neural control engineering for cognitive control. Our path ahead is a compelling opportunity for interdisciplinarians. To enjoy the best that cognitive control and control theory researchers have to offer, we should engage one another to define specific cognitive control problems for applied control theoretic analysis.

**ACKNOWLEDGEMENTS**

The author thanks Joe Kable for the invitation to contribute this review. He additionally thanks Brian Erickson for conversations regarding control theory and systems theory, and Denise Harvey for conversations about the relationship between network control theory and cognitive control. The author acknowledges support from the Office of the Director at the National Institutes of Health through award number DP5-OD021352, the National Institute of Dental and Craniofacial Research through award number R01-DC014960, and a Translational Neuroscience Initiative award through the Perelman School of Medicine.

Kawato, M., Furukawa, K., & Suzuki, R. (1987). A hierarchical neural-network model for control and learning of voluntary movement. Biological cybernetics, 57(3), 169-185.

Karabanov, A., Thielscher, A., & Siebner, H. R. (2016). Transcranial brain stimulation: closing the loop between brain and stimulation. Current opinion in neurology, 29 (4), 397.

Kellert, S. H. (1994). In the wake of chaos: Unpredictable order in dynamical systems. University of Chicago press.

Kerns, J. G., Cohen, J. D., MacDonald, A. W., Cho, R. Y., Stenger, V. A., & Carter, C. S. (2004). Anterior cingulate conflict monitoring and adjustments in control. Science, 303 (5660), 1023-1026.

Kolling, N., Wittmann, M. K., Behrens, T. E., Boorman, E. D., Mars, R. B., & Rushworth, M. F. (2016). Value, search, persistence and model updating in anterior cingulate cortex. Nature neuroscience, 19 (10), 1280.

Korn, H., & Faure, P. (2003). Is there chaos in the brain? ii. experimental evidence and related models. Comptes rendus biologies, 326 (9), 787-840.

Kraus, D., Naros, G., Bauer, R., Khademi, F., Leao, M. T., Ziemann, U., & Gharabaghi, A. (2016). Brain state-dependent transcranial magnetic closed-loop stimulation controlled by sensorimotor desynchronization induces robust increase of corticospinal excitability. Brain Stimulation: Basic, Translational, and Clinical Research in Neuromodulation, 9 (3), 415-424.

Lewis, T. G. (2011). Network science: Theory and applications. John Wiley & Sons.
Lindsay, D. S., & Jacoby, L.L. (1994). Stroop process dissociations: The relationship between facilitation and interference. Journal of Experimental Psychology: Human Perception and Performance, 20 (2), 219.

Liu, Y. J., Li, J., Tong, S., & Chen, C. P. (2016). Neural network control-based adaptive learning design for nonlinear systems with full-state constraints. IEEE transactions on neural networks and learning systems, 27(7), 1562-1571.

Liu, Y.Y., Slotine, J.J., & Barabasi, A.L. (2011). Controllability of complex networks. Nature, 473 (7346), 167-173.

Ljung, L. (1992). System identification toolbox. Math Works.

Ljung, L. (1998). System identification. In Signal analysis and prediction (pp. 163-173). Springer.

Logan, G. D. (1989). Automaticity and cognitive control. Unintended thought, 52-74.
MacDonald, A. W., Cohen, J. D., Stenger, V. A., & Carter, C. S. (2000). Dissociating the role of the dorsolateral prefrontal and anterior cingulate cortex in cognitive control. Science, 288 (5472), 1835-1838.
25